\documentclass[reprint,showpacs,preprintnumbers,nofootinbib,amsmath,amssymb,aps,prd,floatfix]{revtex4-1}
\pdfoutput=1
\usepackage{graphicx}
\usepackage{bm}
\usepackage{subfigure}
\usepackage{url}
\usepackage[hyperindex]{hyperref}
\usepackage{color}
\usepackage[ddmmyy,24hr]{datetime}
\usepackage{bigdelim}
\usepackage{booktabs}
\usepackage{dcolumn}
\usepackage{multirow}
\usepackage{subfigure}
\usepackage{cancel}
\usepackage{stackrel}
\usepackage{paralist}
\usepackage{xspace}
\newcommand{\nua}[1]{\ensuremath{\rlap{\kern-2.5pt\ensuremath{\overset{\scriptscriptstyle(-)}{\phantom{\nu}}}}{\ensuremath{{\nu}_{#1}}}}}

\newcommand\Tstrut{\rule{0pt}{2.6ex}}         
\begin{document}

\title{Improved Determination of the $^{235}\text{U}$ and $^{239}\text{Pu}$ Reactor Antineutrino Cross Sections per Fission}

\author{C. Giunti}
\affiliation{INFN, Sezione di Torino, Via P. Giuria 1, I--10125 Torino, Italy}

\date{9 June 2017}

\begin{abstract}
We present the results of a combined fit of the
reactor antineutrino rates and the Daya Bay
measurement of
$\sigma_{f,235}$ and $\sigma_{f,239}$.
The combined fit
leads to a better determination of the two cross sections per fission:
$\sigma_{f,235}
=
6.29
\pm
0.08$
and
$\sigma_{f,239}
=
4.24
\pm
0.21
$
in units of
$10^{-43} \, \text{cm}^2 / \text{fission}$,
with respective uncertainties of about
$1.2\%$
and
$4.9\%$.
Since
the respective deviations from the theoretical cross sections per fission are
$2.5\sigma$
and
$0.7\sigma$,
we conclude
that,
if the reactor antineutrino anomaly
is not due to active-sterile neutrino oscillations,
it is likely that
it can be solved with a revaluation of the $^{235}\text{U}$ reactor antineutrino flux.
However,
the $^{238}\text{U}$, $^{239}\text{Pu}$, and $^{241}\text{Pu}$ fluxes,
which have larger uncertainties,
could also be significantly different from the theoretical predictions.
\end{abstract}


\maketitle

The flux of electron antineutrinos produced in nuclear reactors
is generated by the $\beta$ decays of the fission products of
$^{235}\text{U}$,
$^{238}\text{U}$,
$^{239}\text{Pu}$, and
$^{241}\text{Pu}$.
The 2011 recalculation
\cite{Mueller:2011nm,Huber:2011wv}
of the four fluxes led to the discovery of the
reactor antineutrino anomaly
\cite{Mention:2011rk},
which is a deficit of the rate of electron antineutrinos measured in several
reactor neutrino experiments.
There are two known possible explanations of the reactor antineutrino anomaly:
1) a miscalculation of one or more of the four electron antineutrino fluxes
\cite{Giunti:2016elf,An:2017osx}
and
2) active-sterile neutrino oscillations
(see Ref.~\cite{Gariazzo:2017fdh} and references therein).
In this paper we consider the first possibility
and we present an improvement of the results presented in
Refs.~\cite{Giunti:2016elf,An:2017osx}
on the determination of the
cross sections per fission
$\sigma_{f,235}$
and
$\sigma_{f,239}$,
which are,
respectively,
the integrals of the products of the
$^{235}\text{U}$ and $^{239}\text{Pu}$
electron antineutrino fluxes
and the detection cross section
[see Eq.~(8) of Ref.~\cite{Mention:2011rk}].

The cross section per fission $\sigma_{f,235}$
of the
$^{235}\text{U}$
electron antineutrino flux
was determined in Ref.~\cite{Giunti:2016elf}
with a fit of the reactor rates by taking into account the different fuel compositions.
Recently the Daya Bay Collaboration presented a
determination of
$\sigma_{f,235}$
and
$\sigma_{f,239}$
obtained by measuring the correlations between
the reactor core fuel evolution and
the changes in the reactor antineutrino flux and energy spectrum
\cite{An:2017osx}.
In this paper we present a combined fit of the
reactor rates
and the Daya Bay measurement of
$\sigma_{f,235}$
and
$\sigma_{f,239}$
which leads to a better determination of both cross sections per fission.

In the analysis of the reactor rates,
we consider the theoretical ratios
\cite{Giunti:2016elf}
\begin{equation}
R_{a}^{\text{th}}
=
\dfrac{\sum_{k} f^{a}_{k} r_{k} \sigma_{f,k}^{\text{SH}}}{\sum_{k} f^{a}_{k} \sigma_{f,k}^{\text{SH}}}
,
\label{ratio}
\end{equation}
where
$f^{a}_{k}$ is the antineutrino flux fraction from the fission of the
isotope with atomic mass $k$
and
the coefficient $r_{k}$
is the corresponding correction of the
theoretical cross section per fission $\sigma_{f,k}^{\text{SH}}$
which is needed to fit the data
($k = 235, 238, 239, 241$,
denotes, respectively, the
$^{235}\text{U}$,
$^{238}\text{U}$,
$^{239}\text{Pu}$,
$^{241}\text{Pu}$
electron antineutrino fluxes).
The theoretical cross sections per fission
$\sigma_{f,k}^{\text{SH}}$
are the Saclay+Huber (SH) \cite{Mention:2011rk,Huber:2011wv}
cross sections per fission
listed in Table~1 of Ref.~\cite{Giunti:2016elf}.
The index $a$ labels the reactor neutrino experiments
listed in Table~1 of Ref.~\cite{Gariazzo:2017fdh}:
Bugey-4 \cite{Declais:1994ma},
Rovno91 \cite{Kuvshinnikov:1990ry},
Bugey-3 \cite{Declais:1995su},
Gosgen \cite{Zacek:1986cu},
ILL \cite{Kwon:1981ua,Hoummada:1995zz},
Krasnoyarsk87 \cite{Vidyakin:1987ue},
Krasnoyarsk94 \cite{Vidyakin:1990iz,Vidyakin:1994ut},
Rovno88 \cite{Afonin:1988gx},
SRP \cite{Greenwood:1996pb},
Nucifer \cite{Boireau:2015dda},
Chooz \cite{Apollonio:2002gd},
Palo Verde \cite{Boehm:2001ik},
Daya Bay \cite{An:2016srz},
RENO \cite{RENO-AAP2016},
and
Double Chooz \cite{DoubleChooz-private-16}.

We analyze the data of the reactor rates
with the least-squares statistic
\begin{align}
\chi^2_{\text{R}}
=
\null & \null
\sum_{a,b}
\left( R_{a}^{\text{th}} - R_{a}^{\text{exp}} \right)
\left( V_{\text{R}}^{-1} \right)_{ab}
\left( R_{b}^{\text{th}} - R_{b}^{\text{exp}} \right)
\nonumber
\\
\null & \null
+
\sum_{k=238,241}
\left( \dfrac{1-r_{k}}{\Delta{r}_{k}} \right)^2
,
\label{chiR}
\end{align}
where
$R_{a}^{\text{exp}}$
are the measured reactor rates listed in Table~1 of Ref.~\cite{Gariazzo:2017fdh}
and
$V_{\text{R}}$ is the covariance matrix constructed with the corresponding uncertainties.
The second term in Eq.~(\ref{chiR})
serves to keep under control the variation of the rates of the minor fissionable isotopes
$^{238}\text{U}$
and
$^{241}\text{Pu}$,
which are not well determined by the fit \cite{Giunti:2016elf}.
We consider
$\Delta{r}_{238}=15\%$
and
$\Delta{r}_{241}=10\%$,
which are significantly larger than the nominal theoretical uncertainties
(respectively,
8.15\% and 2.15\% \cite{Mention:2011rk,Huber:2011wv})
and the 5\% estimate in
Ref.~\cite{Hayes:2016qnu}.

\begin{table}[!t]
\centering
\begin{ruledtabular}
\begin{tabular}{ccccc}
&
SH
&
Reactor Rates
&
Daya Bay
&
Combined
\\
\hline\Tstrut
$\sigma_{f,235}$
&
$6.69 \pm 0.14$
&
$
6.35
\pm
0.09
$
&
$6.17 \pm 0.17$
&
$
6.29
\pm
0.08
$
\\
$\sigma_{f,239}$
&
$4.40 \pm 0.11$
&
$
3.82
\pm
0.43
$
&
$4.27 \pm 0.26$
&
$
4.24
\pm
0.21
$
\end{tabular}
\end{ruledtabular}
\caption{ \label{tab:csf}
Comparison of the theoretical Saclay+Huber (SH)
values of the cross sections per fission
$\sigma_{f,235}$ and $\sigma_{f,239}$
with those
obtained from the fit of the reactor rates,
from the Daya Bay data \cite{An:2017osx},
and from the combined fit.
The units are $10^{-43} \, \text{cm}^2 / \text{fission}$.
}
\end{table}

\begin{figure}[!t]
\includegraphics*[width=\linewidth]{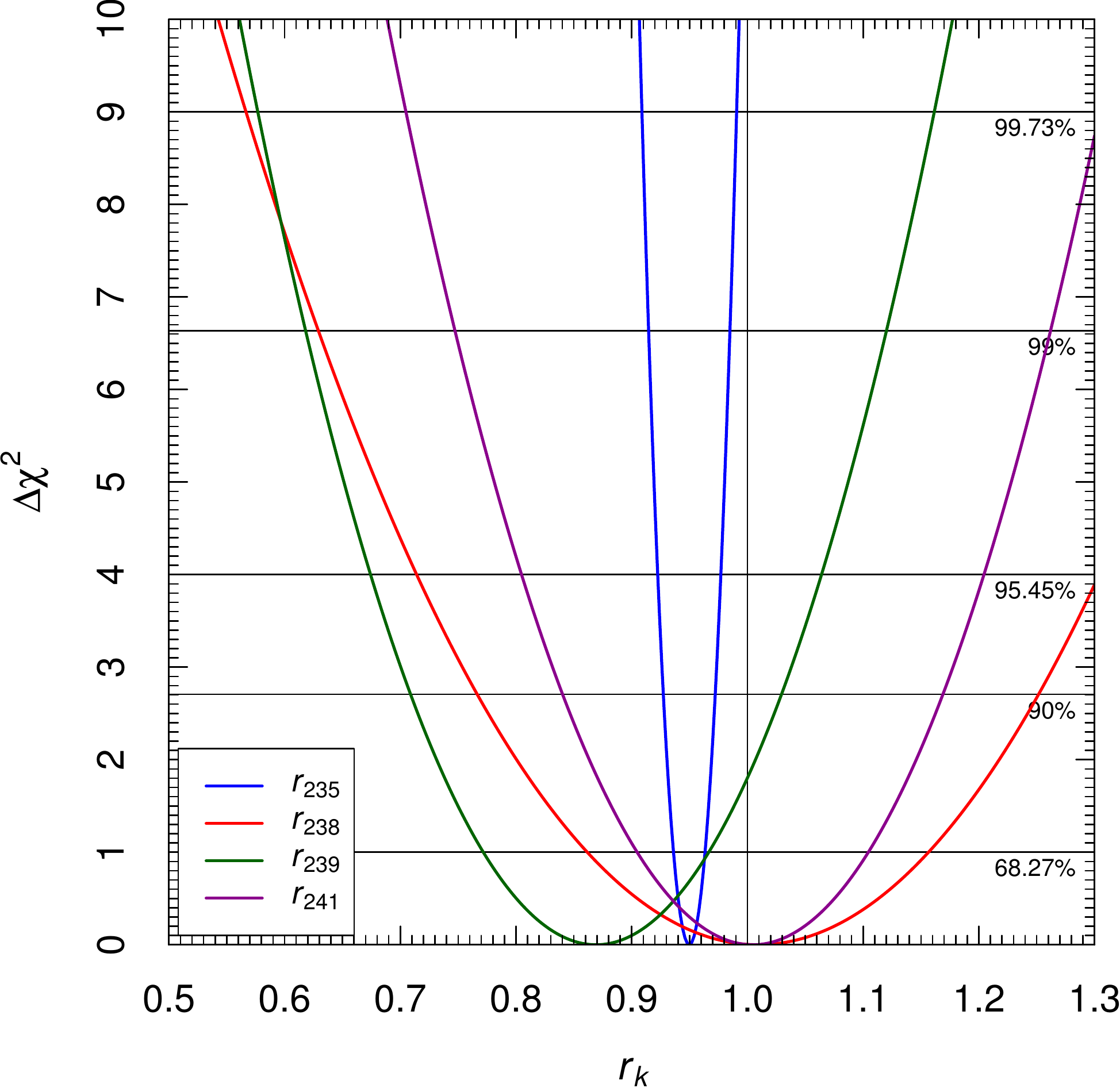}
\caption{ \label{fig:plt1}
Marginal $\Delta\chi^2_{\text{R}} = \chi^2_{\text{R}} - (\chi^2_{\text{R}})_{\text{min}}$
for the coefficients $r_{k}$ of the four antineutrino fluxes
obtained from the fit of the reactor rates.
}
\end{figure}

The fit of the data gives
$(\chi^2_{\text{R}})_{\text{min}} = 17.7$
with
$22$
degrees of freedom,
which correspond to an excellent
$72\%$
goodness of fit.
Figure~\ref{fig:plt1} shows the marginal
$\Delta\chi^2_{\text{R}} = \chi^2_{\text{R}} - (\chi^2_{\text{R}})_{\text{min}}$
for the coefficients $r_{k}$ of the four antineutrino fluxes,
for which we obtain:
\begin{align}
r_{235}
=
\null & \null
0.950
\pm
0.014
,
\label{rat1-235}
\\
r_{238}
=
\null & \null
1.009
\pm
0.147
,
\label{rat1-238}
\\
r_{239}
=
\null & \null
0.869
\pm
0.097
,
\label{rat1-239}
\\
r_{241}
=
\null & \null
1.005
\pm
0.100
.
\label{rat1-241}
\end{align}
These values and Fig.~\ref{fig:plt1}
are different from the corresponding ones in Ref.~\cite{Giunti:2016elf},
because of the different second term in Eq.~(\ref{chiR})
with respect to that in Eq.~(8) of Ref.~\cite{Giunti:2016elf},
which constrained all the $r_{k}$'s.
The best-fit values and uncertainties of
$\sigma_{f,235}$ and $\sigma_{f,239}$
are given in the second column of Table~\ref{tab:csf}.
The value of
$\sigma_{f,235}$
is determined by the fit with a precision of about
$1.4\%$
and differs from the theoretical value
$\sigma_{f,235}^{\text{SH}}$
by
$2.0\sigma$.
This confirms the necessity of a revaluation of the theoretical value of
$\sigma_{f,235}$
found in Ref.~\cite{Giunti:2016elf}.
The value of
$\sigma_{f,239}$
is also determined by the fit, but with the worse precision of about
$11.2\%$,
which renders it compatible with
the theoretical value
$\sigma_{f,239}^{\text{SH}}$
within
$1.3\sigma$.

\begin{figure}[!t]
\includegraphics*[width=\linewidth]{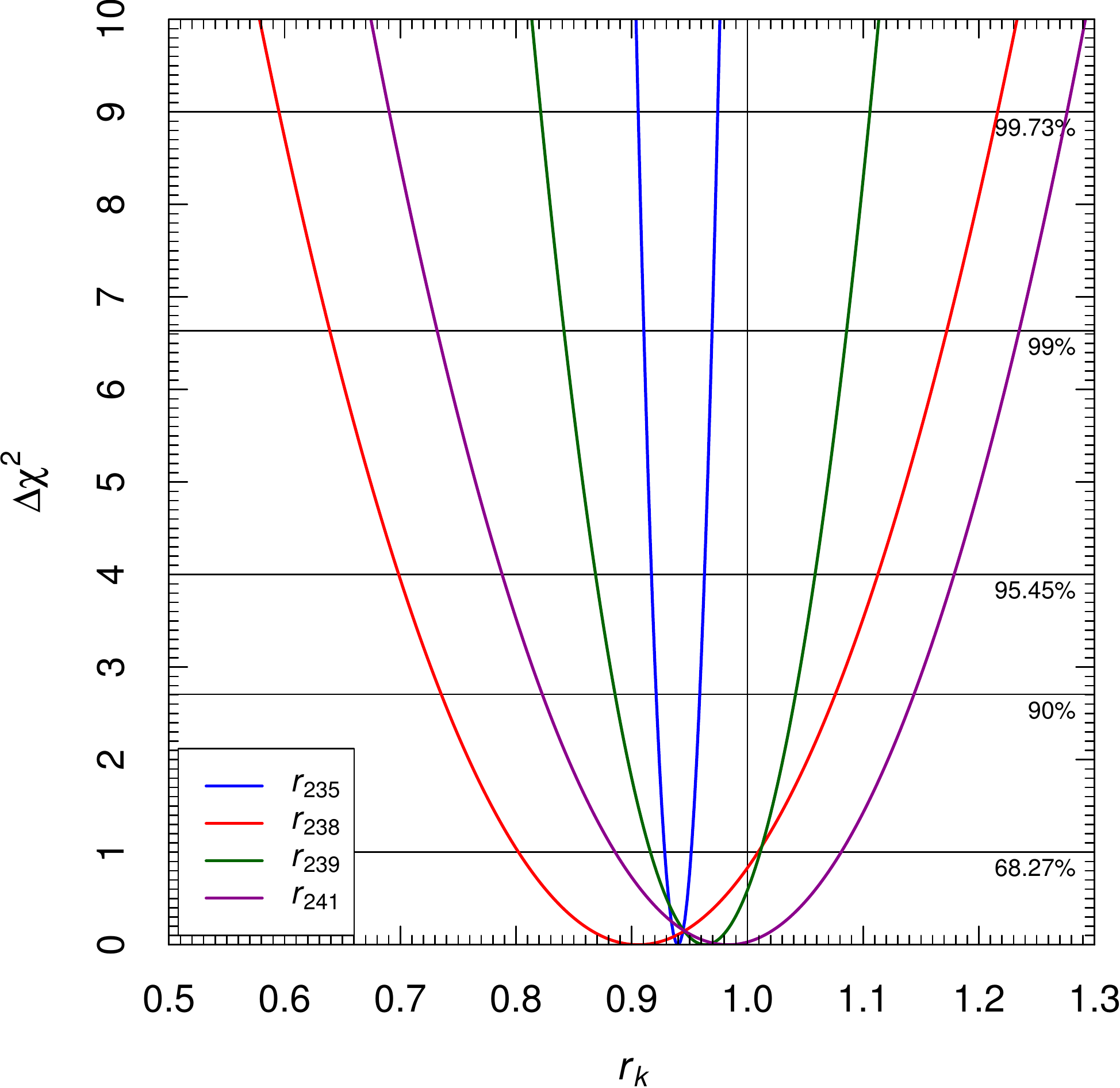}
\caption{ \label{fig:plt2}
Marginal $\Delta\chi^2_{\text{tot}} = \chi^2_{\text{tot}} - (\chi^2_{\text{tot}})_{\text{min}}$
for the coefficients $r_{k}$ of the four antineutrino fluxes
obtained from the fit of the reactor rates and the Daya Bay
measurement of $\sigma_{f,235}$ and $\sigma_{f,239}$ \cite{An:2017osx}.
}
\end{figure}

\begin{figure}[!t]
\includegraphics*[width=\linewidth]{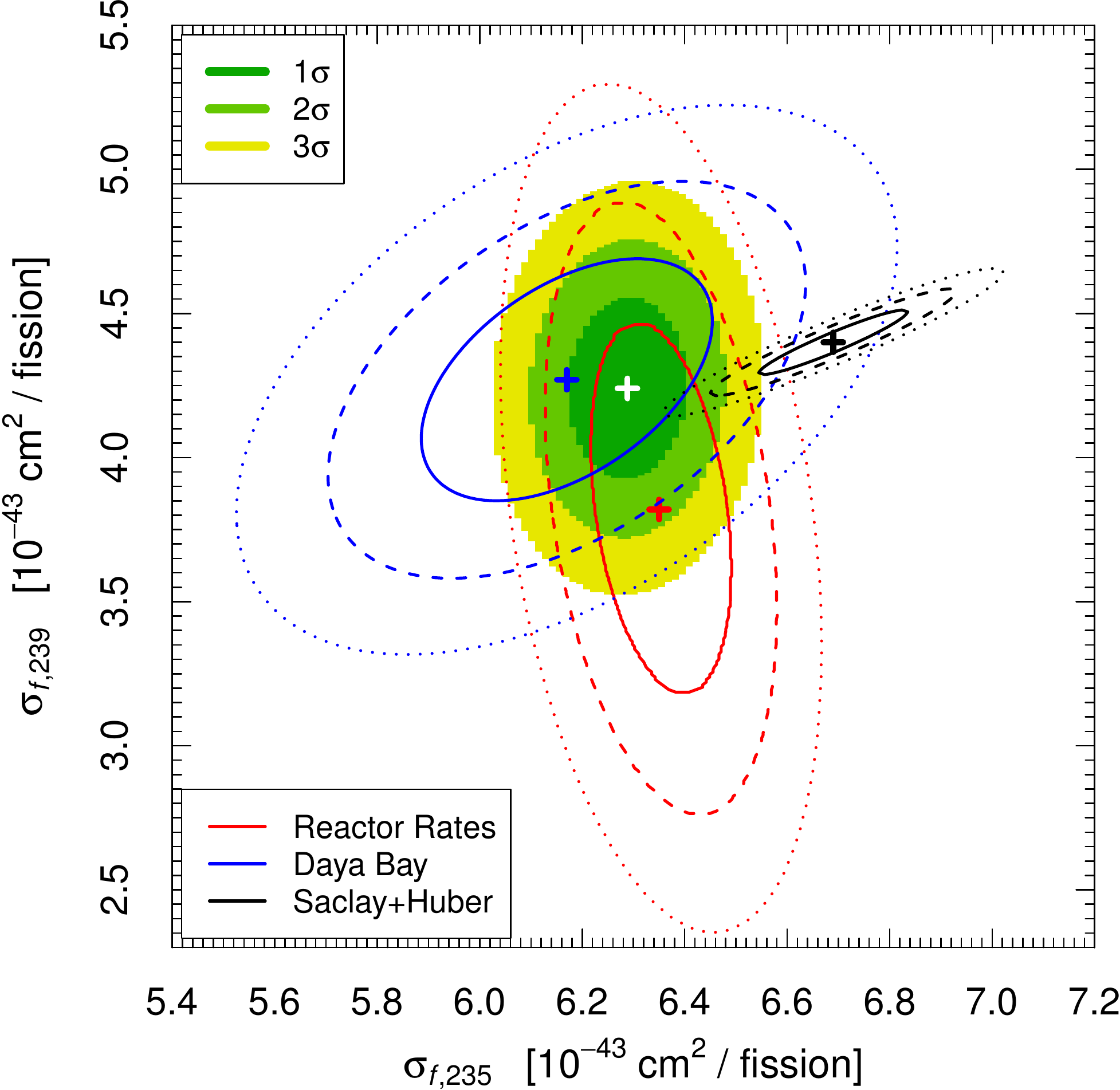}
\caption{ \label{fig:sup-13}
Allowed regions in the
$\sigma_{f,235}$--$\sigma_{f,239}$
plane
obtained from the combined fit of
the reactor rates and the Daya Bay
measurement of $\sigma_{f,235}$ and $\sigma_{f,239}$ \cite{An:2017osx}.
The red, blue and black curves enclose,
respectively,
the allowed regions obtained from the fit of the reactor rates,
the allowed regions corresponding to the Daya Bay
measurement of $\sigma_{f,235}$ and $\sigma_{f,239}$ \cite{An:2017osx},
and the theoretical Saclay+Huber allowed regions
at
$1\sigma$ (solid),
$2\sigma$ (dashed), and
$3\sigma$ (dotted).
The best-fit points are indicated by crosses.
}
\end{figure}

In order to take into account the Daya Bay
measurement of $\sigma_{f,235}$ and $\sigma_{f,239}$ \cite{An:2017osx},
we consider the least-squares statistic
\begin{align}
\null & \null
\chi^2_{\text{tot}}
=
\widetilde\chi^2_{\text{R}}
\nonumber
\\
\null & \null
+
\sum_{k,j=235,239}
\left( \sigma_{f,k}^{\text{th}} - \sigma_{f,k}^{\text{DB}} \right)
\left( V_{\text{DB}}^{-1} \right)_{kj}
\left( \sigma_{f,j}^{\text{th}} - \sigma_{f,j}^{\text{DB}} \right)
,
\label{chi}
\end{align}
where
$\widetilde\chi^2_{\text{R}}$
is given by Eq.~(\ref{chiR})
without considering the Daya Bay rate \cite{An:2016srz},
in order to avoid considering the Daya Bay data twice.
The cross sections per fission
$\sigma_{f,235}^{\text{DB}}$
and
$\sigma_{f,239}^{\text{DB}}$
are those measured in Daya Bay \cite{An:2017osx}
and listed in the third column of Table~\ref{tab:csf}.
We obtained the Daya Bay covariance matrix
$V_{\text{DB}}$
with a Gaussian approximation of the $\chi^2$ distribution
in Fig.~3 of Ref.~\cite{An:2017osx}.
The theoretical cross sections per fission
$\sigma_{f,k}^{\text{th}}$
are given by
\begin{equation}
\sigma_{f,k}^{\text{th}}
=
r_{k} \, \sigma_{f,k}^{\text{SH}}
,
\label{sigmath}
\end{equation}
with the same coefficients $r_{k}$
that are present in the definition of
$R_{a}^{\text{th}}$
in Eq.~(\ref{ratio}).

The minimization of $\chi^2_{\text{tot}}$ gives
$(\chi^2_{\text{tot}})_{\text{min}} = 19.5$
with
$23$
degrees of freedom,
which correspond to a
$67\%$
goodness of fit,
which is practically as good as that obtained in the
analysis of the reactor rates with $\chi^2_{\text{R}}$
in Eq.~(\ref{chiR}).
Figure~\ref{fig:plt2} shows the marginal
$\Delta\chi^2_{\text{tot}} = \chi^2_{\text{tot}} - (\chi^2_{\text{tot}})_{\text{min}}$
for the coefficients $r_{k}$ of the four antineutrino fluxes,
for which we obtain:
\begin{align}
r_{235}
=
\null & \null
0.940
\pm
0.011
,
\label{rat2-235}
\\
r_{238}
=
\null & \null
0.906
\pm
0.103
,
\label{rat2-238}
\\
r_{239}
=
\null & \null
0.964
\pm
0.047
,
\label{rat2-239}
\\
r_{241}
=
\null & \null
0.983
\pm
0.097
.
\label{rat2-241}
\end{align}
The corresponding best-fit values and uncertainties of
$\sigma_{f,235}$ and $\sigma_{f,239}$
are given in the fourth column of Table~\ref{tab:csf}.
The value of
$\sigma_{f,235}$
is determined by the fit with a precision which is slightly better than
that obtained from the fit of the reactor rates,
and significantly better than the precision of the Daya Bay measurement \cite{An:2017osx}.
The combined fit results in a substantial improvement of
the precision of the determination of
$\sigma_{f,239}$
with respect to the fit of the reactor rates alone:
the value of
$\sigma_{f,239}$
is determined with a precision of about
$4.9\%$,
which is also better than that of the Daya Bay measurement \cite{An:2017osx}.
Since the deviation from the theoretical value
$\sigma_{f,239}^{\text{SH}}$
is only of
$0.7\sigma$,
there is no compelling necessity of a revaluation of its
theoretical value.

Figure~\ref{fig:sup-13} shows the correlation between the determinations of
$\sigma_{f,235}$ and $\sigma_{f,239}$.
The values of
$\sigma_{f,235}$ and $\sigma_{f,239}$
obtained from the fit of the reactor rates are slightly anticorrelated,
whereas the Daya Bay values are significantly correlated
and
have a larger uncertainty for $\sigma_{f,235}$
and smaller uncertainty for $\sigma_{f,239}$.
The combined fit results in an allowed region with practically uncorrelated values of
$\sigma_{f,235}$ and $\sigma_{f,239}$
and significantly smaller uncertainties.

The $2.5\sigma$
deviation of $\sigma_{f,235}$ from the theoretical Saclay+Huber
\cite{Mention:2011rk,Huber:2011wv}
cross sections per fission
confirms the indications obtained in Refs.~\cite{Giunti:2016elf,An:2017osx}
that the reactor antineutrino anomaly
is most probably mainly due to the $^{235}\text{U}$ electron antineutrino flux
(if is not due to active-sterile neutrino oscillations).
This possibility
may be connected with a $^{235}\text{U}$ origin of the
5 MeV bump
of the reactor antineutrino spectrum measured
in the
RENO \cite{Seo:2014xei,RENO:2015ksa},
Double Chooz \cite{Abe:2014bwa},
Daya Bay \cite{An:2016srz},
and
NEOS \cite{Ko:2016owz}
experiments,
as indicated by the analysis in Ref.~\cite{Huber:2016xis}
and by the hint of a correlation in the RENO experiment
\cite{RENO-AAP2016}.
The new reactor experiments
PROSPECT \cite{Ashenfelter:2015uxt},
SoLid \cite{Ryder:2015sma}, and
STEREO \cite{Helaine:2016bmc}
which are in preparation for the search of short-baseline neutrino oscillations
with highly enriched $^{235}\text{U}$ research reactors,
will improve the determination
of the $^{235}\text{U}$ electron antineutrino flux.

Since the
$^{238}\text{U}$
and
$^{241}\text{Pu}$
fuel composition in power reactors is small
(see Table~1 of Ref.~\cite{Gariazzo:2017fdh}),
the antineutrino data do not give precise information on the
corresponding cross sections per fission.
From Fig.~\ref{fig:plt2} one can see that
$
r_{238}
=
0.906
\pm
0.103
$
and
$
r_{241}
=
0.983
\pm
0.097
$.
Hence,
there is an indication that
$\sigma_{f,238}$
may be substantially smaller than the theoretical
$\sigma_{f,238}^{\text{SH}}$
value,
but the discrepancy is less than $1\sigma$.
On the other hand,
the fit favors a value of
$\sigma_{f,241}$
close to the theoretical
$\sigma_{f,241}^{\text{SH}}$
value,
but the uncertainty is large.

The calculations of the
$^{235}\text{U}$,
$^{239}\text{Pu}$, and
$^{241}\text{Pu}$
antineutrino fluxes were performed through the inversion
of the corresponding electron spectra measured at ILL in the 80's
\cite{Schreckenbach:1985ep,Hahn:1989zr}.
A possible explanation of the discrepancy between
the calculated and measured values of
$\sigma_{f,235}$
alone could be some unknown systematic error in the measurement of the
$^{235}\text{U}$
electron spectrum
which was not present in the measurements of the
$^{239}\text{Pu}$ and $^{241}\text{Pu}$
electron electron spectra.
It is clear that it would be very important to
check these measurements with new experiments.

In conclusion,
we performed a combined fit of the
reactor antineutrino rates \cite{Giunti:2016elf} and the recent
Daya Bay
measurement of $\sigma_{f,235}$ and $\sigma_{f,239}$ \cite{An:2017osx}.
The combined fit
leads to the better determination of
$\sigma_{f,235}$ and $\sigma_{f,239}$
in Table~\ref{tab:csf},
with respective uncertainties of about
$1.2\%$
and
$4.9\%$.
The respective deviations from the theoretical Saclay+Huber
\cite{Mention:2011rk,Huber:2011wv}
cross sections per fission are
$2.5\sigma$
and
$0.7\sigma$.
Therefore,
we confirm the conclusion already reached in Refs.~\cite{Giunti:2016elf,An:2017osx}
that
the $^{235}\text{U}$ reactor antineutrino flux
is the most probable main contributor to the reactor antineutrino anomaly
\cite{Mention:2011rk}
if the anomaly
is not due to active-sterile neutrino oscillations.
However,
also the
$^{239}\text{Pu}$
flux,
which is constrained by the cross section per fission in Table~\ref{tab:csf},
and
the $^{238}\text{U}$ and $^{241}\text{Pu}$ fluxes,
for which the data do not provide stringent constraints,
could be significantly different from the theoretical predictions.
Let us finally emphasize that the knowledge of the reactor antineutrino fluxes is
useful not only for applications in fundamental physics research,
but also for practical applications as antineutrino monitoring of reactors
(see Refs.~\cite{Hayes:2011ci,Christensen:2014pva,Hayes:2017ymu}).

%

\begin{thebibliography}{37}%
\makeatletter
\providecommand \@ifxundefined [1]{%
\@ifx{#1\undefined}
}%
\providecommand \@ifnum [1]{%
\ifnum #1\expandafter \@firstoftwo
\else \expandafter \@secondoftwo
\fi
}%
\providecommand \@ifx [1]{%
\ifx #1\expandafter \@firstoftwo
\else \expandafter \@secondoftwo
\fi
}%
\providecommand \natexlab [1]{#1}%
\providecommand \enquote [1]{``#1''}%
\providecommand \bibnamefont [1]{#1}%
\providecommand \bibfnamefont [1]{#1}%
\providecommand \citenamefont [1]{#1}%
\providecommand \href@noop [0]{\@secondoftwo}%
\providecommand \href [0]{\begingroup \@sanitize@url \@href}%
\providecommand \@href[1]{\@@startlink{#1}\@@href}%
\providecommand \@@href[1]{\endgroup#1\@@endlink}%
\providecommand \@sanitize@url [0]{\catcode `\\12\catcode `\$12\catcode
`\&12\catcode `\#12\catcode `\^12\catcode `\_12\catcode `\%12\relax}%
\providecommand \@@startlink[1]{}%
\providecommand \@@endlink[0]{}%
\providecommand \url [0]{\begingroup\@sanitize@url \@url }%
\providecommand \@url [1]{\endgroup\@href {#1}{\urlprefix }}%
\providecommand \urlprefix [0]{URL }%
\providecommand \Eprint [0]{\href }%
\providecommand \doibase [0]{http://dx.doi.org/}%
\providecommand \selectlanguage [0]{\@gobble}%
\providecommand \bibinfo [0]{\@secondoftwo}%
\providecommand \bibfield [0]{\@secondoftwo}%
\providecommand \translation [1]{[#1]}%
\providecommand \BibitemOpen [0]{}%
\providecommand \bibitemStop [0]{}%
\providecommand \bibitemNoStop [0]{.\EOS\space}%
\providecommand \EOS [0]{\spacefactor3000\relax}%
\providecommand \BibitemShut [1]{\csname bibitem#1\endcsname}%
\let\auto@bib@innerbib\@empty
\bibitem [{\citenamefont {Mueller}\ \emph {et~al.}(2011)\citenamefont {Mueller}
\emph {et~al.}}]{Mueller:2011nm}%
\BibitemOpen
\bibfield {author} {\bibinfo {author} {\bibfnamefont {T.~A.}\ \bibnamefont
{Mueller}} \emph {et~al.},\ }\href@noop {} {\bibfield {journal} {\bibinfo
{journal} {Phys. Rev.}\ }\textbf {\bibinfo {volume} {C83}},\ \bibinfo {pages}
{054615} (\bibinfo {year} {2011})},\ \Eprint
{http://arxiv.org/abs/arXiv:1101.2663} {arXiv:1101.2663 [hep-ex]}
\BibitemShut {NoStop}%
\bibitem [{\citenamefont {Huber}(2011)}]{Huber:2011wv}%
\BibitemOpen
\bibfield {author} {\bibinfo {author} {\bibfnamefont {P.}~\bibnamefont
{Huber}},\ }\href@noop {} {\bibfield {journal} {\bibinfo {journal} {Phys.
Rev.}\ }\textbf {\bibinfo {volume} {C84}},\ \bibinfo {pages} {024617}
(\bibinfo {year} {2011})},\ \Eprint {http://arxiv.org/abs/arXiv:1106.0687}
{arXiv:1106.0687 [hep-ph]} \BibitemShut {NoStop}%
\bibitem [{\citenamefont {Mention}\ \emph {et~al.}(2011)\citenamefont {Mention}
\emph {et~al.}}]{Mention:2011rk}%
\BibitemOpen
\bibfield {author} {\bibinfo {author} {\bibfnamefont {G.}~\bibnamefont
{Mention}} \emph {et~al.},\ }\href@noop {} {\bibfield {journal} {\bibinfo
{journal} {Phys. Rev.}\ }\textbf {\bibinfo {volume} {D83}},\ \bibinfo {pages}
{073006} (\bibinfo {year} {2011})},\ \Eprint
{http://arxiv.org/abs/arXiv:1101.2755} {arXiv:1101.2755 [hep-ex]}
\BibitemShut {NoStop}%
\bibitem [{\citenamefont {Giunti}(2017)}]{Giunti:2016elf}%
\BibitemOpen
\bibfield {author} {\bibinfo {author} {\bibfnamefont {C.}~\bibnamefont
{Giunti}},\ }\href@noop {} {\bibfield {journal} {\bibinfo {journal}
{Phys.Lett.}\ }\textbf {\bibinfo {volume} {B764}},\ \bibinfo {pages} {145}
(\bibinfo {year} {2017})},\ \Eprint {http://arxiv.org/abs/arXiv:1608.04096}
{arXiv:1608.04096 [hep-ph]} \BibitemShut {NoStop}%
\bibitem [{\citenamefont {An}\ \emph {et~al.}(2017{\natexlab{a}})\citenamefont
{An} \emph {et~al.}}]{An:2017osx}%
\BibitemOpen
\bibfield {author} {\bibinfo {author} {\bibfnamefont {F.~P.}\ \bibnamefont
{An}} \emph {et~al.} (\bibinfo {collaboration} {Daya Bay}),\ }\href@noop {}
{\bibfield {journal} {\bibinfo {journal} {Phys.Rev.Lett.}\ }\textbf
{\bibinfo {volume} {118}},\ \bibinfo {pages} {251801} (\bibinfo {year}
{2017}{\natexlab{a}})},\ \Eprint {http://arxiv.org/abs/arXiv:1704.01082}
{arXiv:1704.01082 [physics]} \BibitemShut {NoStop}%
\bibitem [{\citenamefont {Gariazzo}\ \emph {et~al.}(2017)\citenamefont
{Gariazzo}, \citenamefont {Giunti}, \citenamefont {Laveder},\ and\
\citenamefont {Li}}]{Gariazzo:2017fdh}%
\BibitemOpen
\bibfield {author} {\bibinfo {author} {\bibfnamefont {S.}~\bibnamefont
{Gariazzo}}, \bibinfo {author} {\bibfnamefont {C.}~\bibnamefont {Giunti}},
\bibinfo {author} {\bibfnamefont {M.}~\bibnamefont {Laveder}}, \ and\
\bibinfo {author} {\bibfnamefont {Y.}~\bibnamefont {Li}},\ }\href@noop {}
{\bibfield {journal} {\bibinfo {journal} {JHEP}\ }\textbf {\bibinfo
{volume} {1706}},\ \bibinfo {pages} {135} (\bibinfo {year} {2017})},\ \Eprint
{http://arxiv.org/abs/arXiv:1703.00860} {arXiv:1703.00860 [hep-ph]}
\BibitemShut {NoStop}%
\bibitem [{\citenamefont {Declais}\ \emph {et~al.}(1994)\citenamefont {Declais}
\emph {et~al.}}]{Declais:1994ma}%
\BibitemOpen
\bibfield {author} {\bibinfo {author} {\bibfnamefont {Y.}~\bibnamefont
{Declais}} \emph {et~al.} (\bibinfo {collaboration} {Bugey}),\ }\href@noop {}
{\bibfield {journal} {\bibinfo {journal} {Phys. Lett.}\ }\textbf {\bibinfo
{volume} {B338}},\ \bibinfo {pages} {383} (\bibinfo {year}
{1994})}\BibitemShut {NoStop}%
\bibitem [{\citenamefont {Kuvshinnikov}\ \emph {et~al.}(1991)\citenamefont
{Kuvshinnikov}, \citenamefont {Mikaelyan}, \citenamefont {Nikolaev},
\citenamefont {Skorokhvatov},\ and\ \citenamefont
{Etenko}}]{Kuvshinnikov:1990ry}%
\BibitemOpen
\bibfield {author} {\bibinfo {author} {\bibfnamefont {A.}~\bibnamefont
{Kuvshinnikov}}, \bibinfo {author} {\bibfnamefont {L.}~\bibnamefont
{Mikaelyan}}, \bibinfo {author} {\bibfnamefont {S.}~\bibnamefont {Nikolaev}},
\bibinfo {author} {\bibfnamefont {M.}~\bibnamefont {Skorokhvatov}}, \ and\
\bibinfo {author} {\bibfnamefont {A.}~\bibnamefont {Etenko}},\ }\href@noop {}
{\bibfield {journal} {\bibinfo {journal} {JETP Lett.}\ }\textbf {\bibinfo
{volume} {54}},\ \bibinfo {pages} {253} (\bibinfo {year} {1991})}\BibitemShut
{NoStop}%
\bibitem [{\citenamefont {Achkar}\ \emph {et~al.}(1995)\citenamefont {Achkar}
\emph {et~al.}}]{Declais:1995su}%
\BibitemOpen
\bibfield {author} {\bibinfo {author} {\bibfnamefont {B.}~\bibnamefont
{Achkar}} \emph {et~al.} (\bibinfo {collaboration} {Bugey}),\ }\href@noop {}
{\bibfield {journal} {\bibinfo {journal} {Nucl. Phys.}\ }\textbf {\bibinfo
{volume} {B434}},\ \bibinfo {pages} {503} (\bibinfo {year}
{1995})}\BibitemShut {NoStop}%
\bibitem [{\citenamefont {Zacek}\ \emph {et~al.}(1986)\citenamefont {Zacek}
\emph {et~al.}}]{Zacek:1986cu}%
\BibitemOpen
\bibfield {author} {\bibinfo {author} {\bibfnamefont {G.}~\bibnamefont
{Zacek}} \emph {et~al.} (\bibinfo {collaboration} {CalTech-SIN-TUM}),\
}\href@noop {} {\bibfield {journal} {\bibinfo {journal} {Phys. Rev.}\
}\textbf {\bibinfo {volume} {D34}},\ \bibinfo {pages} {2621} (\bibinfo {year}
{1986})}\BibitemShut {NoStop}%
\bibitem [{\citenamefont {Kwon}\ \emph {et~al.}(1981)\citenamefont {Kwon} \emph
{et~al.}}]{Kwon:1981ua}%
\BibitemOpen
\bibfield {author} {\bibinfo {author} {\bibfnamefont {H.}~\bibnamefont
{Kwon}} \emph {et~al.},\ }\href@noop {} {\bibfield {journal} {\bibinfo
{journal} {Phys. Rev.}\ }\textbf {\bibinfo {volume} {D24}},\ \bibinfo {pages}
{1097} (\bibinfo {year} {1981})}\BibitemShut {NoStop}%
\bibitem [{\citenamefont {Hoummada}\ \emph {et~al.}(1995)\citenamefont
{Hoummada}, \citenamefont {Lazrak~Mikou}, \citenamefont {Bagieu},
\citenamefont {Cavaignac},\ and\ \citenamefont
{Holm~Koang}}]{Hoummada:1995zz}%
\BibitemOpen
\bibfield {author} {\bibinfo {author} {\bibfnamefont {A.}~\bibnamefont
{Hoummada}}, \bibinfo {author} {\bibfnamefont {S.}~\bibnamefont
{Lazrak~Mikou}}, \bibinfo {author} {\bibfnamefont {G.}~\bibnamefont
{Bagieu}}, \bibinfo {author} {\bibfnamefont {J.}~\bibnamefont {Cavaignac}}, \
and\ \bibinfo {author} {\bibfnamefont {D.}~\bibnamefont {Holm~Koang}},\
}\href@noop {} {\bibfield {journal} {\bibinfo {journal} {Applied Radiation
and Isotopes}\ }\textbf {\bibinfo {volume} {46}},\ \bibinfo {pages} {449}
(\bibinfo {year} {1995})}\BibitemShut {NoStop}%
\bibitem [{\citenamefont {Vidyakin}\ \emph {et~al.}(1987)\citenamefont
{Vidyakin} \emph {et~al.}}]{Vidyakin:1987ue}%
\BibitemOpen
\bibfield {author} {\bibinfo {author} {\bibfnamefont {G.~S.}\ \bibnamefont
{Vidyakin}} \emph {et~al.} (\bibinfo {collaboration} {Krasnoyarsk}),\
}\href@noop {} {\bibfield {journal} {\bibinfo {journal} {Sov. Phys. JETP}\
}\textbf {\bibinfo {volume} {66}},\ \bibinfo {pages} {243} (\bibinfo {year}
{1987})}\BibitemShut {NoStop}%
\bibitem [{\citenamefont {Vidyakin}\ \emph {et~al.}(1990)\citenamefont
{Vidyakin} \emph {et~al.}}]{Vidyakin:1990iz}%
\BibitemOpen
\bibfield {author} {\bibinfo {author} {\bibfnamefont {G.~S.}\ \bibnamefont
{Vidyakin}} \emph {et~al.} (\bibinfo {collaboration} {Krasnoyarsk}),\
}\href@noop {} {\bibfield {journal} {\bibinfo {journal} {Sov. Phys. JETP}\
}\textbf {\bibinfo {volume} {71}},\ \bibinfo {pages} {424} (\bibinfo {year}
{1990})}\BibitemShut {NoStop}%
\bibitem [{\citenamefont {Vidyakin}\ \emph {et~al.}(1994)\citenamefont
{Vidyakin} \emph {et~al.}}]{Vidyakin:1994ut}%
\BibitemOpen
\bibfield {author} {\bibinfo {author} {\bibfnamefont {G.~S.}\ \bibnamefont
{Vidyakin}} \emph {et~al.} (\bibinfo {collaboration} {Krasnoyarsk}),\
}\href@noop {} {\bibfield {journal} {\bibinfo {journal} {JETP Lett.}\
}\textbf {\bibinfo {volume} {59}},\ \bibinfo {pages} {390} (\bibinfo {year}
{1994})}\BibitemShut {NoStop}%
\bibitem [{\citenamefont {Afonin}\ \emph {et~al.}(1988)\citenamefont {Afonin}
\emph {et~al.}}]{Afonin:1988gx}%
\BibitemOpen
\bibfield {author} {\bibinfo {author} {\bibfnamefont {A.~I.}\ \bibnamefont
{Afonin}} \emph {et~al.},\ }\href@noop {} {\bibfield {journal} {\bibinfo
{journal} {Sov. Phys. JETP}\ }\textbf {\bibinfo {volume} {67}},\ \bibinfo
{pages} {213} (\bibinfo {year} {1988})}\BibitemShut {NoStop}%
\bibitem [{\citenamefont {Greenwood}\ \emph {et~al.}(1996)\citenamefont
{Greenwood} \emph {et~al.}}]{Greenwood:1996pb}%
\BibitemOpen
\bibfield {author} {\bibinfo {author} {\bibfnamefont {Z.~D.}\ \bibnamefont
{Greenwood}} \emph {et~al.},\ }\href@noop {} {\bibfield {journal} {\bibinfo
{journal} {Phys. Rev.}\ }\textbf {\bibinfo {volume} {D53}},\ \bibinfo {pages}
{6054} (\bibinfo {year} {1996})}\BibitemShut {NoStop}%
\bibitem [{\citenamefont {Boireau}\ \emph {et~al.}(2016)\citenamefont {Boireau}
\emph {et~al.}}]{Boireau:2015dda}%
\BibitemOpen
\bibfield {author} {\bibinfo {author} {\bibfnamefont {G.}~\bibnamefont
{Boireau}} \emph {et~al.} (\bibinfo {collaboration} {NUCIFER}),\ }\href@noop
{} {\bibfield {journal} {\bibinfo {journal} {Phys. Rev.}\ }\textbf
{\bibinfo {volume} {D93}},\ \bibinfo {pages} {112006} (\bibinfo {year}
{2016})},\ \Eprint {http://arxiv.org/abs/arXiv:1509.05610} {arXiv:1509.05610
[physics]} \BibitemShut {NoStop}%
\bibitem [{\citenamefont {Apollonio}\ \emph {et~al.}(2003)\citenamefont
{Apollonio} \emph {et~al.}}]{Apollonio:2002gd}%
\BibitemOpen
\bibfield {author} {\bibinfo {author} {\bibfnamefont {M.}~\bibnamefont
{Apollonio}} \emph {et~al.} (\bibinfo {collaboration} {CHOOZ}),\ }\href@noop
{} {\bibfield {journal} {\bibinfo {journal} {Eur. Phys. J.}\ }\textbf
{\bibinfo {volume} {C27}},\ \bibinfo {pages} {331} (\bibinfo {year}
{2003})},\ \Eprint {http://arxiv.org/abs/hep-ex/0301017} {hep-ex/0301017}
\BibitemShut {NoStop}%
\bibitem [{\citenamefont {Boehm}\ \emph {et~al.}(2001)\citenamefont {Boehm}
\emph {et~al.}}]{Boehm:2001ik}%
\BibitemOpen
\bibfield {author} {\bibinfo {author} {\bibfnamefont {F.}~\bibnamefont
{Boehm}} \emph {et~al.} (\bibinfo {collaboration} {Palo Verde}),\ }\href@noop
{} {\bibfield {journal} {\bibinfo {journal} {Phys. Rev.}\ }\textbf
{\bibinfo {volume} {D64}},\ \bibinfo {pages} {112001} (\bibinfo {year}
{2001})},\ \Eprint {http://arxiv.org/abs/hep-ex/0107009} {hep-ex/0107009}
\BibitemShut {NoStop}%
\bibitem [{\citenamefont {An}\ \emph {et~al.}(2017{\natexlab{b}})\citenamefont
{An} \emph {et~al.}}]{An:2016srz}%
\BibitemOpen
\bibfield {author} {\bibinfo {author} {\bibfnamefont {F.}~\bibnamefont {An}}
\emph {et~al.} (\bibinfo {collaboration} {Daya Bay}),\ }\href@noop {}
{\bibfield {journal} {\bibinfo {journal} {Chin.Phys.}\ }\textbf {\bibinfo
{volume} {C41}},\ \bibinfo {pages} {013002} (\bibinfo {year}
{2017}{\natexlab{b}})},\ \Eprint {http://arxiv.org/abs/arXiv:1607.05378}
{arXiv:1607.05378 [hep-ex]} \BibitemShut {NoStop}%
\bibitem [{\citenamefont {Seo}(2016)}]{RENO-AAP2016}%
\BibitemOpen
\bibfield {author} {\bibinfo {author} {\bibfnamefont {H.}~\bibnamefont
{Seo}},\ }\href@noop {} {\ (\bibinfo {year} {2016})},\ \bibinfo {note} {talk
presented at {AAP 2016, Applied Antineutrino Physics, 1-2 December 2016,
Liverpool, UK}}\BibitemShut {NoStop}%
\bibitem [{Dou()}]{DoubleChooz-private-16}%
\BibitemOpen
\href@noop {} {\ }\bibinfo {note} {{Double Chooz Collaboration, Private
Communication}}\BibitemShut {NoStop}%
\bibitem [{\citenamefont {Hayes}\ and\ \citenamefont
{Vogel}(2016)}]{Hayes:2016qnu}%
\BibitemOpen
\bibfield {author} {\bibinfo {author} {\bibfnamefont {A.~C.}\ \bibnamefont
{Hayes}}\ and\ \bibinfo {author} {\bibfnamefont {P.}~\bibnamefont {Vogel}},\
}\href@noop {} {\bibfield {journal} {\bibinfo {journal}
{Ann.Rev.Nucl.Part.Sci.}\ }\textbf {\bibinfo {volume} {66}},\ \bibinfo
{pages} {219} (\bibinfo {year} {2016})},\ \Eprint
{http://arxiv.org/abs/arXiv:1605.02047} {arXiv:1605.02047 [hep-ph]}
\BibitemShut {NoStop}%
\bibitem [{\citenamefont {Seo}(2015)}]{Seo:2014xei}%
\BibitemOpen
\bibfield {author} {\bibinfo {author} {\bibfnamefont {S.-H.}\ \bibnamefont
{Seo}} (\bibinfo {collaboration} {RENO}),\ }\href@noop {} {\bibfield
{journal} {\bibinfo {journal} {AIP Conf. Proc.}\ }\textbf {\bibinfo {volume}
{1666}},\ \bibinfo {pages} {080002} (\bibinfo {year} {2015})},\ \Eprint
{http://arxiv.org/abs/arXiv:1410.7987} {arXiv:1410.7987 [hep-ex]}
\BibitemShut {NoStop}%
\bibitem [{\citenamefont {Choi}\ \emph {et~al.}(2016)\citenamefont {Choi} \emph
{et~al.}}]{RENO:2015ksa}%
\BibitemOpen
\bibfield {author} {\bibinfo {author} {\bibfnamefont {J.}~\bibnamefont
{Choi}} \emph {et~al.} (\bibinfo {collaboration} {RENO}),\ }\href@noop {}
{\bibfield {journal} {\bibinfo {journal} {Phys. Rev. Lett.}\ }\textbf
{\bibinfo {volume} {116}},\ \bibinfo {pages} {211801} (\bibinfo {year}
{2016})},\ \Eprint {http://arxiv.org/abs/arXiv:1511.05849} {arXiv:1511.05849
[hep-ex]} \BibitemShut {NoStop}%
\bibitem [{\citenamefont {Abe}\ \emph {et~al.}(2014)\citenamefont {Abe} \emph
{et~al.}}]{Abe:2014bwa}%
\BibitemOpen
\bibfield {author} {\bibinfo {author} {\bibfnamefont {Y.}~\bibnamefont
{Abe}} \emph {et~al.} (\bibinfo {collaboration} {Double Chooz}),\ }\href
{\doibase 10.1007/JHEP02(2015)074, 10.1007/JHEP10(2014)086} {\bibfield
{journal} {\bibinfo {journal} {JHEP}\ }\textbf {\bibinfo {volume} {10}},\
\bibinfo {pages} {086} (\bibinfo {year} {2014})},\ \bibinfo {note} {[Erratum:
JHEP 02, 074 (2015)]},\ \Eprint {http://arxiv.org/abs/arXiv:1406.7763}
{arXiv:1406.7763 [hep-ex]} \BibitemShut {NoStop}%
\bibitem [{\citenamefont {Ko}\ \emph {et~al.}(2017)\citenamefont {Ko} \emph
{et~al.}}]{Ko:2016owz}%
\BibitemOpen
\bibfield {author} {\bibinfo {author} {\bibfnamefont {Y.}~\bibnamefont {Ko}}
\emph {et~al.} (\bibinfo {collaboration} {NEOS}),\ }\href@noop {} {\bibfield
{journal} {\bibinfo {journal} {Phys.Rev.Lett.}\ }\textbf {\bibinfo {volume}
{118}},\ \bibinfo {pages} {121802} (\bibinfo {year} {2017})},\ \Eprint
{http://arxiv.org/abs/arXiv:1610.05134} {arXiv:1610.05134 [hep-ex]}
\BibitemShut {NoStop}%
\bibitem [{\citenamefont {Huber}(2017)}]{Huber:2016xis}%
\BibitemOpen
\bibfield {author} {\bibinfo {author} {\bibfnamefont {P.}~\bibnamefont
{Huber}},\ }\href@noop {} {\bibfield {journal} {\bibinfo {journal} {Phys.
Rev. Lett.}\ }\textbf {\bibinfo {volume} {118}},\ \bibinfo {pages} {042502}
(\bibinfo {year} {2017})},\ \Eprint {http://arxiv.org/abs/arXiv:1609.03910}
{arXiv:1609.03910 [hep-ph]} \BibitemShut {NoStop}%
\bibitem [{\citenamefont {Ashenfelter}\ \emph {et~al.}(2016)\citenamefont
{Ashenfelter} \emph {et~al.}}]{Ashenfelter:2015uxt}%
\BibitemOpen
\bibfield {author} {\bibinfo {author} {\bibfnamefont {J.}~\bibnamefont
{Ashenfelter}} \emph {et~al.} (\bibinfo {collaboration} {PROSPECT}),\
}\href@noop {} {\bibfield {journal} {\bibinfo {journal} {J. Phys.}\
}\textbf {\bibinfo {volume} {G43}},\ \bibinfo {pages} {113001} (\bibinfo
{year} {2016})},\ \Eprint {http://arxiv.org/abs/arXiv:1512.02202}
{arXiv:1512.02202 [physics]} \BibitemShut {NoStop}%
\bibitem [{\citenamefont {Ryder}(2015)}]{Ryder:2015sma}%
\BibitemOpen
\bibfield {author} {\bibinfo {author} {\bibfnamefont {N.}~\bibnamefont
{Ryder}} (\bibinfo {collaboration} {SoLid}),\ }\href@noop {} {\bibfield
{journal} {\bibinfo {journal} {PoS}\ }\textbf {\bibinfo {volume}
{EPS-HEP2015}},\ \bibinfo {pages} {071} (\bibinfo {year} {2015})},\ \Eprint
{http://arxiv.org/abs/arXiv:1510.07835} {arXiv:1510.07835 [hep-ex]}
\BibitemShut {NoStop}%
\bibitem [{\citenamefont {Helaine}()}]{Helaine:2016bmc}%
\BibitemOpen
\bibfield {author} {\bibinfo {author} {\bibfnamefont {V.}~\bibnamefont
{Helaine}} (\bibinfo {collaboration} {STEREO}),\ }\href@noop {} {\ }\Eprint
{http://arxiv.org/abs/arXiv:1604.08877} {arXiv:1604.08877 [physics.ins-det]}
\BibitemShut {NoStop}%
\bibitem [{\citenamefont {Schreckenbach}\ \emph {et~al.}(1985)\citenamefont
{Schreckenbach}, \citenamefont {Colvin}, \citenamefont {Gelletly},\ and\
\citenamefont {Von~Feilitzsch}}]{Schreckenbach:1985ep}%
\BibitemOpen
\bibfield {author} {\bibinfo {author} {\bibfnamefont {K.}~\bibnamefont
{Schreckenbach}}, \bibinfo {author} {\bibfnamefont {G.}~\bibnamefont
{Colvin}}, \bibinfo {author} {\bibfnamefont {W.}~\bibnamefont {Gelletly}}, \
and\ \bibinfo {author} {\bibfnamefont {F.}~\bibnamefont {Von~Feilitzsch}},\
}\href@noop {} {\bibfield {journal} {\bibinfo {journal} {Phys. Lett.}\
}\textbf {\bibinfo {volume} {B160}},\ \bibinfo {pages} {325} (\bibinfo {year}
{1985})}\BibitemShut {NoStop}%
\bibitem [{\citenamefont {Hahn}\ \emph {et~al.}(1989)\citenamefont {Hahn} \emph
{et~al.}}]{Hahn:1989zr}%
\BibitemOpen
\bibfield {author} {\bibinfo {author} {\bibfnamefont {A.~A.}\ \bibnamefont
{Hahn}} \emph {et~al.},\ }\href@noop {} {\bibfield {journal} {\bibinfo
{journal} {Phys. Lett.}\ }\textbf {\bibinfo {volume} {B218}},\ \bibinfo
{pages} {365} (\bibinfo {year} {1989})}\BibitemShut {NoStop}%
\bibitem [{\citenamefont {Hayes}\ \emph {et~al.}(2012)\citenamefont {Hayes},
\citenamefont {Trellue}, \citenamefont {Nieto},\ and\ \citenamefont
{WIlson}}]{Hayes:2011ci}%
\BibitemOpen
\bibfield {author} {\bibinfo {author} {\bibfnamefont {A.~C.}\ \bibnamefont
{Hayes}}, \bibinfo {author} {\bibfnamefont {H.~R.}\ \bibnamefont {Trellue}},
\bibinfo {author} {\bibfnamefont {M.~M.}\ \bibnamefont {Nieto}}, \ and\
\bibinfo {author} {\bibfnamefont {W.~B.}\ \bibnamefont {WIlson}},\ }\href
{\doibase 10.1103/PhysRevC.85.024617} {\bibfield {journal} {\bibinfo
{journal} {Phys. Rev.}\ }\textbf {\bibinfo {volume} {C85}},\ \bibinfo {pages}
{024617} (\bibinfo {year} {2012})},\ \Eprint
{http://arxiv.org/abs/arXiv:1110.0534} {arXiv:1110.0534 [nucl-th]}
\BibitemShut {NoStop}%
\bibitem [{\citenamefont {Christensen}\ \emph {et~al.}(2014)\citenamefont
{Christensen}, \citenamefont {Huber}, \citenamefont {Jaffke},\ and\
\citenamefont {Shea}}]{Christensen:2014pva}%
\BibitemOpen
\bibfield {author} {\bibinfo {author} {\bibfnamefont {E.}~\bibnamefont
{Christensen}}, \bibinfo {author} {\bibfnamefont {P.}~\bibnamefont {Huber}},
\bibinfo {author} {\bibfnamefont {P.}~\bibnamefont {Jaffke}}, \ and\ \bibinfo
{author} {\bibfnamefont {T.}~\bibnamefont {Shea}},\ }\href@noop {} {\bibfield
{journal} {\bibinfo {journal} {Phys. Rev. Lett.}\ }\textbf {\bibinfo
{volume} {113}},\ \bibinfo {pages} {042503} (\bibinfo {year} {2014})},\
\Eprint {http://arxiv.org/abs/arXiv:1403.7065} {arXiv:1403.7065 [physics]}
\BibitemShut {NoStop}%
\bibitem [{\citenamefont {Hayes}(2017)}]{Hayes:2017ymu}%
\BibitemOpen
\bibfield {author} {\bibinfo {author} {\bibfnamefont {A.~C.}\ \bibnamefont
{Hayes}},\ }\href {\doibase 10.1088/1361-6633/80/2/026301} {\bibfield
{journal} {\bibinfo {journal} {Rept. Prog. Phys.}\ }\textbf {\bibinfo
{volume} {80}},\ \bibinfo {pages} {026301} (\bibinfo {year} {2017})},\
\Eprint {http://arxiv.org/abs/arXiv:1701.02756} {arXiv:1701.02756 [nucl-th]}
\BibitemShut {NoStop}%
\end{thebibliography}

\end{document}